\documentclass[a4paper,12pt]{article}
\pdfoutput=1
\usepackage{jcappub} 
\usepackage{newtxtext,newtxmath}
\usepackage{color,xcolor}
\usepackage{lineno}
\usepackage{graphicx}	
\usepackage{latexsym,amsmath,amssymb}
\usepackage{multicol}
\usepackage{tikz}   
\usetikzlibrary{arrows}
\usepackage{subfigure,float}
\usepackage{dblfloatfix}
\usepackage{newtxmath}
\usepackage{subcaption}
\usepackage[export]{adjustbox}
\usetikzlibrary{shapes.geometric, arrows}
\graphicspath{{Figures/}} 
\usepackage{multicol}
\usepackage[normalem]{ulem}
\usepackage{parskip} 
\usepackage{fixltx2e} 
\usepackage{soul,ulem}

\def \HI{{\sc Hi}}

\arxivnumber{2407.17573} 
\title{Ionospheric effect on the synthetic Epoch of Reionization observations with the SKA1-Low}

\author[a,1]{Samit Kumar Pal,\note{Corresponding author.}}\emailAdd{palsamitkumar@gmail.com}
\author[a]{Abhirup Datta,}\emailAdd{abhirup.datta@iiti.ac.in}
\author[b,a]{and Aishrila Mazumder}\emailAdd{aishri0208@gmail.com}
\affiliation[a]{Department of Astronomy, Astrophysics \& Space Engineering, Indian Institute of Technology Indore, Indore 453552, India}
\affiliation[b]{Jodrell Bank Centre for Astrophysics, Department of Physics and Astronomy, The University of Manchester, Manchester M13 9PL, UK}
\date{\today}
\abstract{The redshifted $21$\,cm signal of neutral hydrogen can be used as a direct probe of the intergalactic medium during Cosmic Dawn\,(CD) and Epoch of Reionization\,(EoR). However, detecting this inherently weak signal has numerous challenges. The major ones include accurate foreground removal from low-frequency radio observations and systematics arising from instrumental effects. The Earth's ionosphere poses a major obstacle at these low radio frequencies. Thus, a systematic study of ionospheric effects on these sensitive low-frequency observations is critical, given that the construction of the Square Kilometre Array (SKA1-Low) is in full progress. We use the end-to-end pipeline, called \textsc{21cmE2E}, to study the effect of time-varying ionospheric corruption on the $21$\,cm power spectrum recovery. We use two models: a) a catalogue-based model focused on source position shift due to the refractive effect of the ionosphere and b) a realistic ionospheric condition generated using Kolmogorov's turbulence model. We assess the effect of the imperfections thus generated on the extraction of \HI\ $21$\,cm signal power spectrum. Our study shows that beyond ``median ionospheric offset" ($\theta_{\text{MIO}} \lesssim 0.1''$), the $21$\,cm signal from the EoR is unaffected by residual ionospheric effects. Our study emphasizes the need for the development of efficient ionospheric calibration algorithms for the upcoming SKA1-Low observations to extract the \HI\ $21$\,cm power spectra from the CD/EoR.}


\begin{document}
 \maketitle
 \flushbottom
\section{Introduction}
\label{introduction}
One of the main aims of the upcoming Square Kilometer Array (SKA1-Low) radio interferometer is to observe the Cosmic Dawn (CD) and Epoch of Reionization (EoR) era. The CD/EoR is a critical period for the structure formation in the early universe. It marks the transition from neutral hydrogen (\HI) dominated intergalactic medium (IGM) to an ionized one. It occurred when ultraviolet (UV) radiation of the first luminous objects, such as the first stars and galaxies, started stripping electrons from these abundant \HI\ atoms. Many complementary probes used hydrogen, the most abundant baryonic element in the early universe, to study these epochs. Observations of the cosmic microwave background (CMB) and its interaction with free electrons (Thompson scattering \cite{komatsu2011} ) constrain the reionization history \cite{planck2018}. Additionally, studies of the luminosity function and clustering properties of  Ly$\alpha$ emitters \citep{Ouchi2010,ota2017} and high-redshift quasars emitting Ly$\alpha$ radiation \citep{Fan2003, Goto2012, Barnett2017} provide further insights. This combined evidence suggests that reionization is nearing completion at a redshift of $z\sim 6$ \cite{planck2018}. The \HI\ $21$\,cm line, originates when an electron in the ground state undergoes a spin flip from a parallel to an antiparallel state. This helps us trace the entire reionization history and thus will answer many questions related to EoR science \citep{Fulanetto2006, Tirth2009, pritchard2012}.\\
In the past decade, radio interferometers like the Murchison Widefield Array \citep[MWA, ][]{Barry2019, Yoshiura2021, Kolopanis2023}, the LOw Frequency ARray \citep[LOFAR, ][]{Mertens2020}), Giant Meterwave Radio Telescope \citep[GMRT, ][]{Paciga2013}, Precision Array to Probe Epoch of Reionization \citep[PAPER, ][]{Kolopanis2019} and Hydrogen Epoch of Reionization Array \citep[HERA, ][]{Deboer2017,hera2022, HERA_2023} focused on detecting these signal from the early epoch of the universe. While the \HI\ $21$\,cm signal is the desired observable, it is not the only emission detected. Astrophysical foreground emission, like galactic synchrotron radiation and extragalactic free-free emission, also contributes to the overall measured signal. The desired EoR signal exhibits statistical isotropy and a fluctuating spectrum, whereas the foreground signals exhibit a spectrally smooth spectrum and are thereby structureless. In Fourier space, the spectrally smooth foreground emission is confined to the wedge-shaped region and is localized by the horizon line. A wedge is a definite result of the chromaticity of the instrument. The region expected to be free from foreground contamination in Fourier space is known as the EoR window. Therefore, for extracting the \HI\ $21$\,cm signal, the wedge-shaped structure is avoided. This approach is known as the foreground avoidance technique \citep{Datta_2010, Vedantham2012ApJ...745..176V, Thyagarajan2015ApJ...804...14T, Thyagarajan2015ApJ...807L..28T}. Another method involves subtracting the foreground using either a parametric or non-parametric model, known as the foreground removal technique \citep{ Mertens2018MNRAS.478.3640M, Hothi2021MNRAS.500.2264H, Kern2021}. Precise instrument calibration is essential for accurate foreground modeling and subtraction
\citep{Offringa2015PASA, Barry2016MNRAS, Trott_Wayth2016, Ewall-wice2017MNRAS, Patil2017ApJ...838...65P, Dillon2018MNRAS.477.5670D, Kern2019ApJ...884..105K, Mazumder2022}. There are other contaminations, for example the instrument model \citep{deLera2017MNRAS.469.2662D, Trott2017MNRAS.470..455T, Joseph2018AJ....156..285J, Li2018ApJ...863..170L}, incomplete sky model \citep{Barry2016MNRAS}, Earth's ionosphere \citep{jordan2017, Trott2018ApJ, Helmboldt2020}, and radio frequency interference (RFI) \cite{Wilensky2019}.

Over the past decade, extensive studies have been conducted to understand and quantify the residual excess noise. This excess noise is often attributed to systematics, which primarily constrains the upper limits of the CD/EoR power spectrum measurements. The potential sources of such systematics are residual ionospheric phase errors \citep{Kariuki2022, Gan2023}, inaccurate modelling of the complex antenna beam shapes \citep{Chokshi2024}, and unflagged low-level RFI \citep{Wilensky2019, Wilensky2023}. It is essential to quantify each of these errors, which will introduce the excess noise in the power spectrum estimation from SKA1-Low observations. This study discusses the residual direction-dependent errors (DDEs) caused by inaccuracies in calibrating the ionosphere effects. The Earth's ionosphere, coupled with a bright foreground, plays a major role in the systematic effects. Ionospheric refraction shifts the incident radio waves from the cosmic sources, thereby leading to changes in their positions in the image plane. It occurs due to transverse variation of the density of electrons (total electron content, i.e., TEC) column along the line-of-sight, aligning with the direction of the gradient. This effect varies with change in direction \citep{Lonsdale2005}. The relationship between the $\Vec{\nabla}\text{TEC}$ and the resulting angular offset ${\delta}\Vec{\theta}$ is described by Equation \ref{equ:tecvsangel}:
\begin{equation}
    \Vec{\nabla}\text{TEC} = 1.20\times10^{-4}\left(\frac{\nu}{100\text{ MHz}}\right)^2\left(\frac{{\delta}\Vec{\theta}}{1 ''}\right) \text{TECU km}^{-1}
    \label{equ:tecvsangel}
\end{equation}
This effect primarily depends on the observing frequency (which, in our case, dominates other effects) and the spatial distribution of the TEC in the ionosphere.

Recent work by the MWA and LOFAR has developed methods to calibrate for the ionospheric phase errors over longer time intervals \citep{intema2009, Roja_Dodson_2018, degasperin2018, Mertens2018MNRAS.478.3640M, Tasse_2021, Kariuki2022}. However, there is a need for ionospheric calibration at shorter time intervals matching with ionospheric variability. Previous work done by the MWA EoR experiments corrected the ionospheric phase errors by flagging time bins with bad ionospheric conditions \citep{Trott2018ApJ}. Chege et al. \citep{Kariuki2022} demonstrated that ionospheric refractive effects significantly contribute to excess variance and updating the model using interpolated ionospheric phase screens, but this approach did not yield improved results. Vedantham and Koopmans et al. \citep{Vedantham2015, Vedantham_2016} analytically investigated the expected variance of measured visibilities and the impact of scintillation noise on power spectra considering the simple ionosphere model. Brackenhoff et al. \citep{Brackenhoff_2024} focused on how the transfer of direction-dependent gain calibration errors from longer baselines to short baselines impacts the final power spectrum estimation in LOFAR simulation. However, the accuracy of these calibration techniques is limited by the signal-to-noise ratio (SNR) and the number of directions or sources considered during the calibration process. The residual direction-dependent errors (DDEs) or the residual ionospheric offsets after calibration of the effects of the ionosphere hinder the detection of the target cosmological signal. In this paper, we study ionospheric corruption on the shorter baselines in realistic CD/EoR observations using SKA1-Low, focusing on the time-varying nature of the ionosphere. Here, we investigate the accuracy in ionospheric calibration and the residual ionospheric offsets that will not affect the $21$\,cm PS estimation from SKA1-Low observations. We also assume that other calibration effects, such as direction-independent gain errors, are perfectly corrected and do not contribute to the residual errors in our analysis. For the first time, we present a simulation of ionospheric phase errors to emphasize the critical need for developing efficient ionospheric calibration algorithms for the upcoming SKA1-Low observation to estimate the $21$\,cm PS precisely.

This paper is organized as follows: Section~\ref{sec:syn} details the setup of our simulation framework for synthetic SKA1-Low observations. Section~\ref{imageplane} discusses the impact of ionospheric corruption on extracting the \HI\ $21$\,cm signal from the image plane for different ionospheric scenarios. Section~\ref{ps_estimation} presents the results of PS estimation and the conclusions drawn for EoR. Finally, Section~\ref{sec:summary} summarizes our key findings from our analysis. The best-fitted cosmological parameters from the Planck 2018 results \cite{planck2018} were used throughout this study, with the following details: $\Omega_M= 0.31, \Omega_{\Lambda} =0.68, \sigma_{8}=0.811, H_0=67.36$\,\text{km~s}$^{-1}$~Mpc$^{-1}$.

\section{Simulations}
\label{sec:syn}
This section focuses on generating synthetic observations using ionospheric models. Here, we have used the end-to-end pipeline, \textsc{21cmE2E} outlined in \cite{Mazumder2022}. We use $21$\,cm signal model and T-RECS catalogue as a sky model. In order to study the impact of the DDEs on the shorter baselines in realistic CD/EoR observations using SKA1-Low, we generate mock data with and without residual ionospheric phase errors. Figure~\ref{e2e} shows a schematic block diagram of the \textsc{21cmE2E}\footnote{\url{https://gitlab.com/samit-pal/21cme2e.git}} pipeline. The observation parameters are listed in Table~\ref{tab:obs_para}. The following subsections briefly describe each parameter in the input block. The simulations have been performed on the \textsc{OSKAR}\footnote{\url{https://github.com/OxfordSKA/OSKAR/releases}} package, and Common Astronomy Software Application (\textsc{CASA}\footnote{\url{https://casaguides.nrao.edu/index.php?title=Main_Page}} \cite{McMullin_2007}) was used to construct the synthetic map from model visibility data. \textsc{OSKAR} uses the radio interferometer measurement equation \cite{Hamaker996} to generate full stroke visibility data. In this work, a four-hour observation period ($\pm 2$ HA) tracks the sky with the phase centre pointed at $\alpha = 15h00m00s$ and $\delta = -30^{\circ}00m00s$ using a $120$\,second integration time and bandwidth of $8$\,MHz at a centre frequency of $142$\,MHz ($z\sim 9$) with a channel separation of $125$\,kHz ($64$\,channels). 
\begin{figure}
\begin{center}
\tikzstyle{startstop} = [rectangle, rounded corners, minimum width=3cm, minimum height=1cm,text centered, draw=black , fill=blue!15]
\tikzstyle{io} = [trapezium, trapezium left angle=60, trapezium right angle=120, minimum width=3cm, minimum height=1cm,text width=3cm,, draw=black , fill=blue!15]

\tikzstyle{connector} = [draw, -latex']
\begin{tikzpicture}[node distance=2cm]
\node (start) [startstop] {Foreground Model + Model 21-cm Signal};
\node (data1) [startstop] at (0,-2) {Telescope Model \& Observational Parameters};
\path [connector] (start) -- (data1);
\draw[->] (0,-2.5) -- (0,-4);
\draw (-4,-4) -- (0,-4);
\draw (-4,-4) -- (-4,-6);
\draw (4,-4) -- (0,-4);
\draw (4,-4) -- (4,-6);
\node (data2) [startstop] at (-4,-6) {Without Ionospheric Model};
\node (data3) [startstop] at (4,-6) {With Ionospheric Model};
\node (data4) [startstop] at (-4,-8) {Model Visibility};
\path [connector] (data2) -- (data4);
\node (data5) [startstop] at (4,-8) {Corrupt Visibility};
\path [connector] (data3) -- (data5);
\node[draw=none] at (0,-5)  {Simulation (\textsc{OSKAR})};
\node (data6) [startstop] at  (0,-10) {Residual Visibility (Corrupted - Model)};
\path [connector] (data4) -- (data6);
\path [connector] (data5) -- (data6);
\node[draw=none] at (0, -9) {\sc uvsub};
\node (data7) [startstop] at (-4,-12) {Imaging};
\node (data8) [startstop] at (4,-12) {Power Spectrum (2D \& 1D)};
\path [connector] (data6) -- (data7);
\path [connector] (data6) -- (data8);
\end{tikzpicture}
\caption{Schematic diagram of the \textsc{21cmE2E} pipeline.}
\label{e2e}
\end{center}
\end{figure}

\begin{table}
    \centering
     \caption{An overview of the observational parameters used in the simulations.}
    \label{tab:obs_para}
    \begin{tabular}{lcccr}
    \hline\hline
    & \\
    Parameter &&  Value \\
           & & \\
    \hline\hline
    Central frequency      & &142\,MHz (z$\sim 9$)\\
    Bandwidth              & & 8\,MHz\\
    Spectral Resolution         &  & 125\,kHz\\
    Field of view (FoV)    & &$4^{\circ}$\\
    Number of array elements($N_a$)  && 296\\
    Maximum baseline       & &$\sim 2000$\,m\\
    Synthesized beam       & &$\sim 2.5'$\\
    Polarization            & & Stokes I\\
    No. of snapshots        && 120 \\
    Integration time per snapshot && $2$\,minutes\\
    Phase Center(J2000)     && RA, DEC= 5\,h, $-30\,^{\circ}$\\
    Effective collective area ($A_{\text{eff}}$) & & $962$\,m$^2$ \\
    Core area of an array ($A_{\text{core}}$) & & $12.57$\,km$^2$ \\
    Total collecting area of the & & \\
    array ($A_{\text{col}}$) & & $N_{a}A_{\text{eff}}$ \\
    \hline\hline
    \end{tabular}
   
 \end{table}

\subsection{EoR model}
The differential brightness temperature ($\delta T_b$) of the \HI\ $21$\,cm signal during EoR, can be expressed as: 
\begin{equation}
\label{deltbeq}
    \delta{T_{b}} \approx 27x_{\text{HI}}(1+\delta)\left(\frac{1+z}{10}\right)^{\frac{1}{2}}\left(1-\frac{T_{\text{CMB}}(z)}{T_S}\right)\left(\frac{\Omega_{b}}{0.044}\frac{h}{0.7}\right)
    \left(\frac{\Omega_m}{0.27}\right)^{-\frac{1}{2}}\text{mK}
\end{equation}
where $ x_{\rm HI} $ is the neutral fraction of hydrogen, $\delta$ is the density fluctuation, $H$ is the evolving Hubble constant, $T_{\text{CMB}}(z)$ is the  CMB temperature at a redshift of $z$ and $T_S$ is the spin temperature of the two states of hydrogen \citep{Fulanetto2006}. We use a simulated \HI\ $21$\,cm fiducial lightcone cube generated from the semi-numerical simulation, {\sc 21cmfast} \citep{Mesinger2011, Murray2020}. For our analysis, we used the existing simulated \HI\ $21$\,cm maps from Mazumder et al. \cite{Mazumder2022}. The simulated light cone covers a comoving sky-plane area of $500\times500$\,$h^{-1}$Mpc$^2$, a redshift range of $7.0\leq z \leq 12.0$ and has a grid size of $232\times232\times562$. A specific cuboid region of size $232\times232\times64$ was extracted for analysis. This cuboid was then converted from comoving Mpc to angular-frequency (WCS) coordinate for the input of the \textsc{21cmE2E} pipeline. 

\subsection{Foreground model}
\label{foreground}
A sky model with high resolution in both angular and frequency space is essential to precisely estimate the foregrounds observable by SKA-Low. Following the methodology outlined in \cite{Mazumder2022}, we employ the point source foreground model from the Tiered Radio Extragalactic Continuum Simulation (T-RECS) \cite{bonaldi2019}. T-RECS is a state-of-the-art simulation of the radio sky in continuum, covering a frequency range from 150\,MHz to 20\,GHz. Its model includes active galactic nuclei (AGNs) and star-forming galaxies (SFGs), representing most radio galaxy populations. These simulations are based on recent observational data, ensuring realistic cosmological evolution of luminosity functions, number counts in total intensity and polarization, and clustering properties. This model uses a flux range of $0.6$\,Jy to $3.1$\,mJy at $150$\,MHz. Fluxes were then converted to their corresponding values at $142$\,MHz using a spectral index ($\alpha$) of $-0.8$, which follows the power law relationship, $S_{\nu}\propto \nu^{-\alpha}$. Our sky model comprises of $2522$\,compact sources within a $4$\,degree$^2$. The compact point source modelling allows for a more accurate determination of residual ionospheric offsets. Thus, we excluded the diffuse foreground model and focused solely on point sources only.

\subsection{Telescope model}\label{telescope}
\begin{figure}
    \centering
    \includegraphics[width=0.8\linewidth]{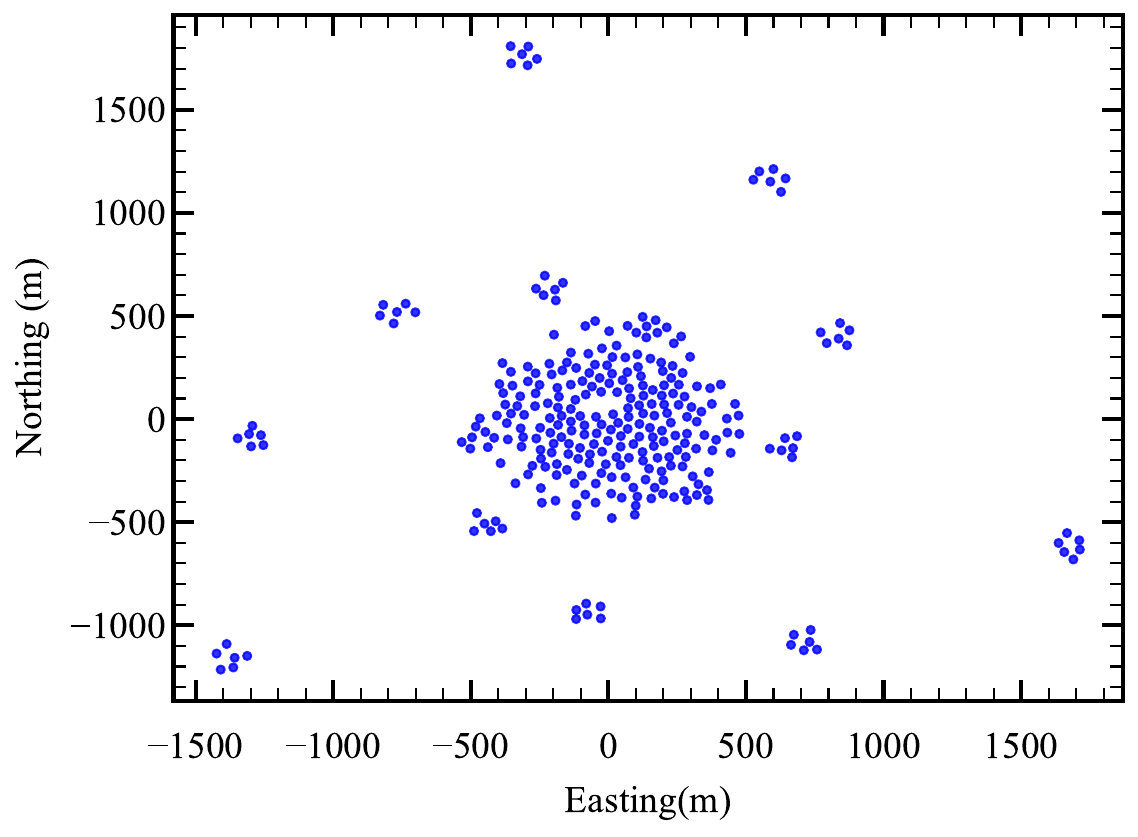}
    \caption{The array assembly 4 (AA4) configuration of SKA1-Low within a radius of $2$\,km around the central core. This array consists of $296$ stations.}
    \label{fig:telescope}
\end{figure}

The highly sensitive next-generation telescope SKA1-Low will make tomographic maps of the \HI\ $21$\,cm signal and precision signal PS measurements. The SKA will consist of two separate arrays: a low-frequency array in Western Australia operating between $50-350$\,MHz, and a mid-frequency array in South Africa operating between $350$\,MHz and $14$\,GHz. CD/EoR observations are done at frequencies $\lesssim$200\,MHz, and will thus use the SKA1-Low. SKA1-Low is being built at the Murchison Radio-astronomy Observatory (MRO) site in Western Australia, where the MWA is currently located. The initial configuration of the SKA1-Low will have $512$ stations of $40$\,m diameter, each with $256$ dipole antennas, dispersed in a dense core and spiral arms with a maximum baseline length of $\sim 70$\,km \cite{ska_design}. The impact of different ionospheric conditions on SKA1-Low will be similar to that of the MWA telescope. This study focuses on the compact core with a maximum baseline distance of $2$\,km from the centre station. We consider the array assembly 4 (AA4) configuration of SKA1-Low within a radius of 2\,km from the array centre \citep{SKAO_telescope}. The compact configuration of the baseline is crucial for EoR observations as the \HI\ $21$\,cm signal presents itself as a diffuse background with its predominant power concentrated on large angular scales or small $k$-scales. Consequently, shorter baselines mostly contribute to the sensitivity of the instrument to the \HI\ signal from EoR. The expected telescope configuration of the SKA1-Low is illustrated in Figure~\ref{fig:telescope}. A detailed study of the ionospheric effect on the longer baselines and other systematics is beyond the scope of this work and will be the subject of a future study.

\subsection{Ionospheric models}
\label{frameworks}
This work aims to demonstrate the impact of the residual direction-dependent errors (DDEs), after calibrating the ionosphere effects using ionospheric calibration techniques. This study shows how these residual errors translate to foreground leakage on the residual visibilities used in the final power spectrum estimation. Ionospheric refraction causes an apparent shift of point sources from their original position. We investigate how the residual DDEs or the residual ionospheric offsets arising from varying levels of ionospheric activity (from extremely active to quiet) affect $k$-modes (spatial scale) in cosmological PS. To quantify the impact of contamination of residual DDEs on PS estimation, we present two approaches for simulating ionospheric effects. The first method leverages a catalogue-based model to introduce residual refraction shifts in point sources based on the previous MWA observations. The second approach employs Kolmogorov's turbulence model to create a more realistic representation of ionospheric conditions reflecting the ambient ionospheric activity at the MRO location. In the second approach, we assess the robustness of the calibration techniques to calibrate the effects of the ionosphere. It is important to note that we have not applied any mitigation techniques, as the goal is to estimate the accuracy of any such techniques that will allow us to detect the CD/EoR signal. The Kolmogorov phase screen does inherently introduce higher-order spatial effects. These effects cannot be adequately modelled by a linear gradient phase screen. However, these higher-order effects are more prominent for interferometric arrays that have longer baselines. Therefore, for baselines within $\sim 2$\,km, the impact of the higher-order effects is expected to be not so dominant and is beyond the scope of this paper. We describe each of these models in detail in the following subsections. The refractive shift and Kolmogorov's turbulence model are referred to as Cases I and II, respectively. Using these frameworks, we can assess how ionospheric contamination manifests itself within CD/EoR science, separating it from other observational effects.

\begin{figure}[b]
	\includegraphics[width=\linewidth]{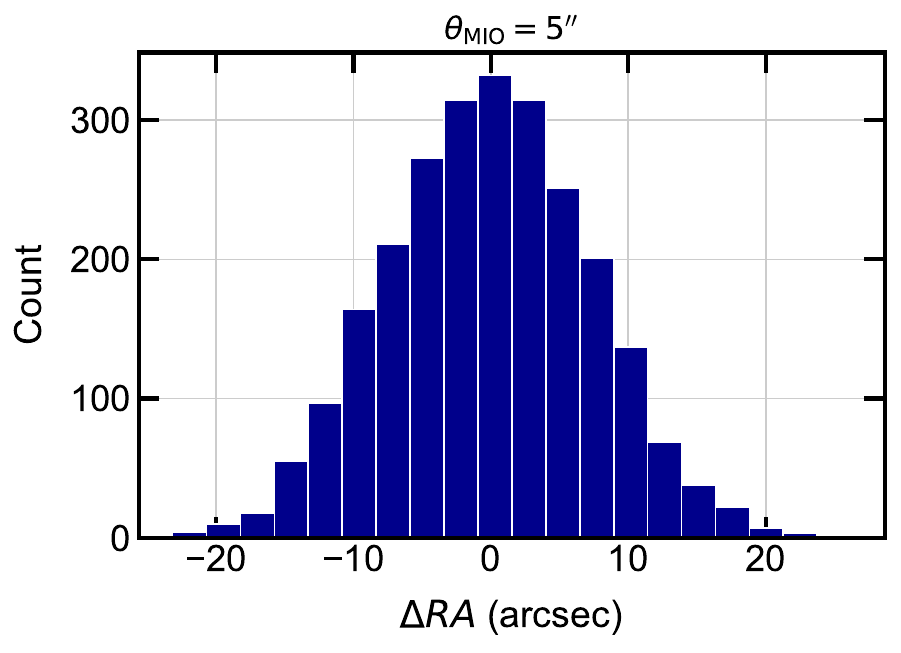}
    \caption{Distribution of angular shifts in the sources of the catalogue with $\theta_{\text{MIO}}=5.0''$, simulated at 142\,MHz for a single snapshot.}
    \label{time_vs_ra}
\end{figure}

\subsubsection{Case I}
\label{position_shift}
The inhomogeneity in density of the plasma of the ionosphere introduces additional phase shifts in electromagnetic waves that vary with time, frequency, and position. In radio interferometry, these phase shifts manifest as apparent shifts in the position of the source in the image plane. To accurately simulate the effect of the ionosphere, we need knowledge about ionospheric conditions at the location of the SKA1-Low telescope. This telescope is being constructed at the MRO site, where the MWA currently operates as a precursor to SKA1-Low. According to the Fourier shift theorem, a position shift in the image plane corresponds to a phase shift in the Fourier domain (i.e., u-v domain). We used a Gaussian distribution to simulate shifts in apparent source positions. This apparent position shift mimics the residual ionospheric offsets after calibration of the effects of the ionosphere.

Previous MWA-based research by Jordan et al.\cite{jordan2017} characterized the ionospheric activity in the MRO based on the position offset of known sources within the field of view of observations. Such position offsets are then characterized by the metric median ionospheric offset (MIO). Furthermore, Jordan et al. \cite{jordan2017} defined quiet ionospheric conditions for the MWA EoR experiment based on a threshold for MIO of $0.15'$ at $200$\,MHz. For this study, we have adapted this MIO-based model for the ionosphere. We have used drawn residual ionospheric offset distributions for different MIOs corresponding to ionospheric activities ranging between active and quiet conditions. In order to simulate the effect of residual ionospheric phase shift, we added the apparent position shift to each source's right ascension (RA) in the T-RECS catalogue following the above MIO-based offset distribution.  These residual offsets can be present in either one of RA or DEC or both in a point source model. For simplicity, we added the residual offsets to the RA of the sources.\\
We have modelled a dynamic ionosphere, updated every $2$\,minutes. This cadence aligns with the time averaging used in the GaLactic and Extragalactic All-Sky MWA (GLEAM) Survey \cite{Wayth2015}. Furthermore, We assume that the source offset exhibits oscillatory behaviour with a period of 4\,hours, effectively mimicking a sinusoidal wave. This results in a temporal evolution of the source residual offsets against a minor travelling ionospheric disturbance (TID). A histogram of the distribution of angular shifts (RA shifts) in the sources of the T-RECS catalogue with $\theta_{\rm MIO} = 5''$ for one such snapshot is shown in Figure~\ref{time_vs_ra}.

\subsubsection{Case II}
\label{phase_screen_kolmo}
In the previous section, the dynamic ionosphere is modelled in terms of a single TID. However, in reality, the ionosphere is a far more complex 3D structure that evolves in time in all directions. Hence, in this model, we incorporate a thin, two-dimensional phase screen \citep{Ratcliffe1956RPPh...19..188R} as the ionospheric model. We have used a phase screen that changes in time using Kolmogorov's turbulence model. Mevius et al. \cite{Mevius2016} reported the value of the spectral index is $\beta = 3.89\pm 0.1$ at the LOFAR site. However, for our simulations, we used the pure Kolmogorov turbulence value of $\beta =11/3$.
In order to model the time-evolving phase screen, we first generated a frozen two-dimensional phase screen at any timestamp $t$, accounting for both inner and outer scales in the spatial phase function. The power spectrum of the spatial phase function is given by,
\begin{equation}
    \left|\Phi (\Vec{k})\right|^2 \propto \left[k^2+ \left(\frac{1}{L_0}\right)^2\right]^{-\beta/2}\text{exp}\left(-\frac{k^2}{2/l_0^2} \right) ;   1/L_0 \ll  k \ll 1/l_0
\end{equation}
where $l_0$ and $L_0$ are the inner and outer scales of the turbulence, respectively, and $\Vec{k}$ is the spatial frequency.
To simulate the statistical behaviour of the new phase screen, we added a linear function to the phase of the complex spectrum of the previous phase screen. This was done by adjusting the speed of the travelling ionospheric disturbance. As a result, the turbulence propagates through space at a certain speed $V_{\text{TID}}$, generating the time-evolving phase screen. This formulation of a time-evolving phase screen is described in Glindemann et al. \citep{Glindemann_1993}. Then the new complex spectrum $\Phi_{\rm new}(\Vec{k})$ becomes
\begin{equation}
     \Phi_{\rm new}(\Vec{k}) = \Phi_{\rm old}(\Vec{k})\text{exp}\left(2\pi i \Vec{k} N_s/N\right)   
\end{equation}
where $N_s$ is the number of pixels by which the phase screen needs to shift in the direction of coordinate $\Vec{k}$ and $N^2$ is the size of the ionospheric layer. The number of pixels is calculated from the ratio between the spatial displacement due to TIDs and the spatial resolution of the pixel in the phase screen. The process is repeated for each new phase screen. Each new state depends only on the preceding phase screen.

\begin{figure}
    \centering
    \includegraphics[width=\linewidth]{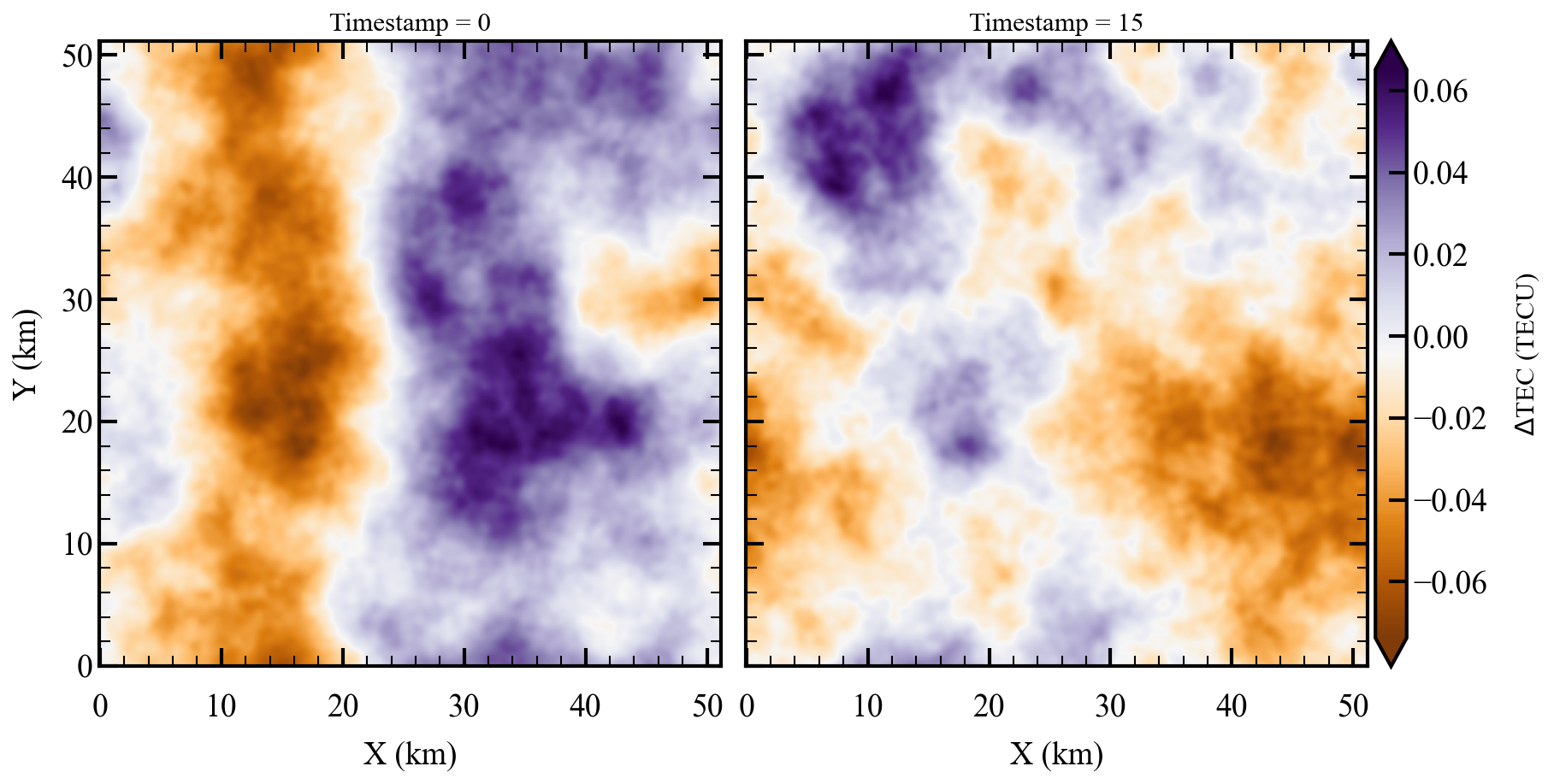}
    \caption{Two slices of a simulated Kolmogorov's turbulence TEC screen were generated using the FFT algorithm with dimensions  $51$\,km$\times51$\,km with $\theta_{\text{MIO}}=5.0''$. At the $400$\,km altitude, this ionospheric screen covered the sources outside the primary FoV.}
    \label{fig:kolmo_screen}
\end{figure}

An understanding of the ionospheric structure is crucial for setting the size of turbulent elements in general models. However, a comprehensive investigation of the finite scale at the MRO site is beyond the scope of this study. Ionospheric irregularities often occur at the equator and high latitudes, whereas mid-latitude locations such as SKA1-Low experience lower activity levels \cite{Fejer1980}. GLEAM observations show that the angular scales of the ionospheric structure are typical $100$\,s of km in the MRO \cite{Helmboldt2020}. Therefore, we chose an outer scale size of $100$\,km for our simulations, consistent with the general estimate provided by GLEAM observations. The SKA1-Low telescope has a field of view (FoV) of $\sim 4^{\circ}$, and we assume the ionosphere is located at an altitude of $400$\,km \cite{jordan2017} above the Earth's atmosphere. This altitude corresponds to a projected ionospheric layer with dimensions of $30$\,km on each side to model the spatial variations of TEC within the field of view. For this study, we simulated a larger ionospheric region of $51\times51$\,km$^2$. We used a travelling ionospheric disturbance speed of $75$\,km/h aligned with the wave vector directions to simulate the turbulent and transient motion of the TID. The dynamic ionosphere, updated every $2$\,minutes on the telescope-projected beam, which is consistent with Case I. Figure~\ref{fig:kolmo_screen} shows two slices of a simulated TEC screen at different timestamps. 

\subsection{Synthetic observations}
\label{sec:method}
The visibility, which encodes the spatial coherence information of the radio sky, is mathematically described by the equation:
\begin{equation}
   V(\boldsymbol{u},\nu) = \int_{\text{sky}} A(\boldsymbol{\hat{s}},\nu)I(\boldsymbol{\hat{s}},\nu)e^{-2i\pi\left(\phi+\phi^{'}\right)} d\Omega
   \label{eq:vis}
\end{equation}
where $\boldsymbol{u}$ is the baseline vector, $I(\boldsymbol{\hat{s}},\nu)$ \& $A(\boldsymbol{\hat{s}},\nu)$ is the specific intensity, and the antenna beam pattern, both of which are functions of frequency $\nu$. The additional phase shift $\phi^{'}$ accounts for the source-position offset in the image domain. The components of the unit vector $\boldsymbol{\hat{s}}$  $(l, m, n)$ are the direction cosines towards the east, north \& zenith, respectively, with $n=\sqrt{1-l^2-m^2 }$ and $d\Omega= dl~dm/\sqrt{1-l^2-m^2}$. In this work, primary beam correction is not considered, i.e., $A(\nu)$ is set to $1$.\\
We simulate a model sky without any bias due to corruption, which is the true or model sky ($V^{\text{model}}_{\rm ij}$). Then we use the inaccurate sky models to determine the effect of the residual ionospheric phase errors. Utilizing the aforementioned ionospheric frameworks \ref{frameworks}, we corrupted visibilities, denoted as $V^{\text{corrupt}}_{\rm ij}$. Simulations are done using the settings described in \autoref{sec:syn}. In both cases, we generated multiple sets of apparent sky models with MIO values of $0.1''$, $0.2''$, $1'', 5''$, and $30''$. These MIO values quantify the inaccuracies of residual DDEs after calibration of the effects of the ionosphere. These residual DDEs exhibit variations arising from fluctuating levels of ionospheric activity, ranging from highly active conditions ($\theta_{\rm MIO} = 30''$) to extremely quiet ($\theta_{\rm MIO} = 0.1''$) conditions. For Case I, we added simulated residual offset distributions with the corresponding MIO values to the actual source positions in RA. In Case II, we scaled the turbulence levels to match the specified MIOs, representing the post-ionospheric calibration accuracy. We quantify the robustness of the calibration technique by adding residual DDEs for each source and station within the \textsc{OSKAR} simulations. In order to validate the impact of refractive phase errors on the model visibility dataset, we employ a constant phase gradient, see Appendix~\ref{constant_phase_screen}.

Now to remove extragalactic point sources from the data, we subtracted the true sky model from both sets of simulated ionospheric contaminated sky models in the visibility domain. This subtraction results in the residual visibility, expressed as:
\begin{equation}
    V^{\text{residual}}_{\rm ij} = V^{\text{corrupt}}_{\rm ij} - V^{\text{model}}_{\rm ij}
\end{equation}
The observed visibility is denoted by $V^{\text{corrupt}}_{\rm ij}$ (the label ``corrupt" denotes the actual observations obtained from the ionospheric effect), and $V^{\text{model}}_{\rm ij}$ is the true visibility without any bias due to corruption. We utilize the residual visibility, as employed in our previous study \cite{Mazumder2022}, to assess the impact on image plane performance and PS estimation. The instrumental noise for SKA1-Low was not added to this simulation. The following sections discuss the formalism used for the image plane effect and PS estimation.

\section{Ionospheric effect on imaging}
\label{imageplane}
In this section, we present the impact of residual ionospheric offsets on the extraction of \HI\ $21$\,cm signal in the image plane. In addition, we analyze the distribution of residual ionospheric offsets after calibrating the ionospheric phase shift.

\subsection{Offset distribution}
\label{offset_distribustion}
We investigate the variation in residual ionospheric offsets in source positions. In order to quantify the residual ionospheric offset distribution, we simulated visibilities including the sky model with and without residual ionospheric effect. The synthetic map was generated using \textsc{CASA}, using Briggs weighting with robust parameter set to $-1$. We extract sources from the synthetic map using {\sc aegean} \footnote{\url{https://github.com/PaulHancock/Aegean}} \cite{aegean_2} source finder tool. Catalogues thus generated from {\sc aegean} have angular positions, flux densities and their associated errors, which are used for studying the impact in the image domain. We select compact sources having high SNR, i.e. flux densities greater than $5$ times the RMS noise in that image ($5\sigma$). The residual offset for each coordinate has been set as: 
\begin{equation}
    \Delta\alpha = \alpha_{\text{ionosphere off}} - \alpha_{\text{ionosphere on}}
    \label{delta_ra}
\end{equation}
\begin{equation}
    \Delta\delta = \delta_{\text{ionosphere off}}  - \delta_{\text{ionosphere on}}
    \label{delta_dec}
\end{equation}
The magnitude of the residual ionospheric offset is calculated as:
\begin{equation}
    \Delta \theta = \sqrt{ \left(\Delta\alpha\right)^2 + \left(\Delta\delta\right)^2 }
    \label{fig:delta_theta}
\end{equation}

\begin{figure*}
    \begin{multicols}{2}
    \includegraphics[ width=\linewidth]{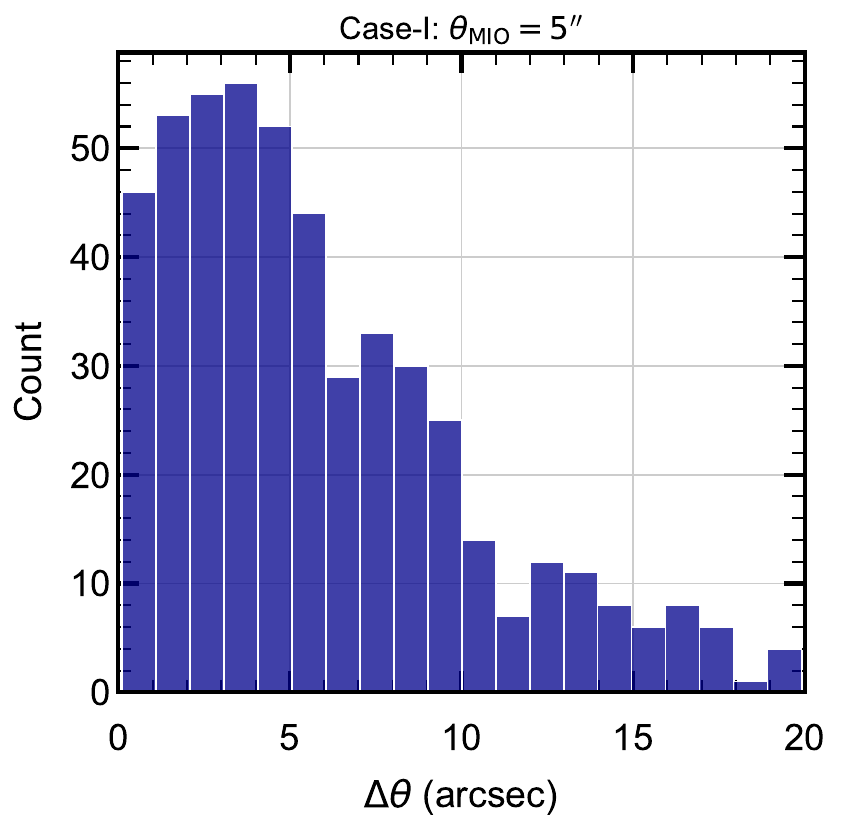}\par
    \includegraphics[width =\linewidth]{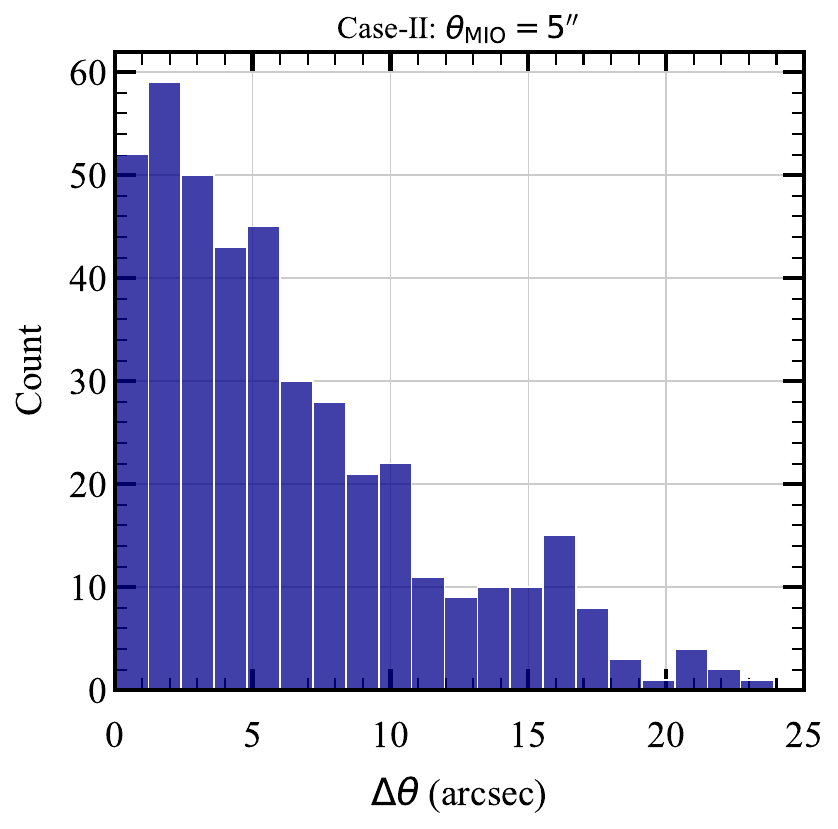}\par
    \end{multicols}
    \caption{Histogram of the distribution of residual ionospheric offsets of selected point sources with MIO of $5''$ for \texttt{(left)} Case I and \texttt{(right)} Case II. Case II has more residual offset outliers than Case I. This significant difference in outlier behaviour translates to vastly different sky models, affecting the subtraction of foreground models and varying tolerance levels in the extraction of the \HI\ signal.}
    \label{fig:offset_dist}
\end{figure*}
Figure~\ref{fig:offset_dist} shows the histogram of the residual ionospheric offset distribution of the selected point sources for both cases with the MIO of $5.0''$. We measured the apparent position shift by calculating the angular displacement of the selected sources from their original position. These shifts vary with time as they pass through a time-varying ionosphere. We estimated the residual ionospheric offset of each source by its time-averaged apparent position across the entire integration time. The residual of the median ionospheric offset ($\theta_{\rm MIO}$) of the selected sources serves as a quantifying metric. This metric signifies the inaccuracy of the ionospheric calibration techniques. Case II exhibits a larger residual offset outlier compared to Case I at the equivalent MIO value. This significant difference in outlier behaviour translates to vastly different sky models, affecting the subtraction of foreground models. Incomplete foreground removal degrades the dynamic range of the EoR signal, further hindering the detection of the target signal. The severity of these effects correlates with source SNR. Lower SNR sources experience a larger apparent offset due to the residual ionospheric errors, leading to a significantly different sky model than high SNR sources. Therefore, this unique contamination in each case poses a significant challenge for extracting the target signal in both PS and image domains.

\subsection{Analysis method}
\label{method_imageplane}
One of the key science projects for SKA1-Low is to make the tomographic images of the \HI\ $21$\,cm signal with a SNR $\gtrapprox 1$ \cite{Mellema:2015IS} at scales from several arcminutes to several degrees. One can use various image-based statistical tools, such as the Minkowski functional \cite{Kaphtia2021} and the Largest Cluster statistic \citep{Bag2018, Dasgupta_2023}, to extract maximum information from the target signal. The residual DDEs lead to the imperfect removal of extragalactic point sources that hinder the detection of the target signal. Therefore, it is essential to assess how the residual ionospheric errors affect the extraction of \HI\  $21$\,cm signal from the image plane.

One of the primary aims of this study is to investigate the degree of tolerance level of the residual ionospheric offsets or the accuracy in ionospheric calibration that will not obscure the target signal from SKA1-Low observations. The effect of Earth's ionosphere significantly limits the dynamic range of the image. The residual DDEs increase the effective RMS noise of the image generated from the residual visibilities. This DDEs limit the achievable dynamic range, regardless of the source flux density. The RMS noise increases as the position of the source increases from the phase centred in the residual image. The RMS noise characteristic can change depending on what kind of gridding scheme is used. In order to estimate the RMS noise from the residual visibility, we generated a dirty image using \textsc{CASA} \texttt{tclean} task using the natural weighting scheme. The RMS noise limit was then estimated within a region near the phase centre of this image. We use RMS noise against the MIO as a quantifying metric in the image plane to analyze the detectability of the \HI\ $21$\,cm signal. However, to achieve detection of the \HI\ $21$\,cm signal, the RMS should be below the signal level. The signal at $142$\,MHz has a peak flux of $\sim 0.8$\,$\mu$Jy~beam$^{-1}$. Thus, for a residual value beyond this, the target signal becomes obscured. We also measured the dynamic range (DR) to assess image plane performance. The dynamic range is the ratio of the highest peak flux and the RMS noise in an unsourced region in the image. Perley et al. \cite{Perley_1999} calculated an equation for DR for $N$ antennas for $M$ independent successive snapshots and is given by,
\begin{equation}
    \text{DR} = \frac{\sqrt{M}N}{\phi_0}
    \label{dr_perley}
\end{equation}
where $\phi_0$ is the time-independent phase error (in radians) for all the baselines.\\
In order to detect the faint EoR signal, we must suppress the instrumental noise level of the instrument, which reflects the sensitivity of the instrument. The instrumental noise of the radio interferometers is estimated from the radiometer equation:
\begin{equation}
    \sigma_N = \left(\frac{2 k_B}{A_{\text{eff}}N_{\text{ant}}}\right)\left(\frac{T_{\text{sys}}}{\sqrt{n_p\Delta\nu t_{int}}}\right)
    \label{equ:thermal}
\end{equation}
where $\sigma_N$ denotes the thermal noise of the instrument, $\Delta \nu$ denotes the channel width, $t_{\text{int}}$ denotes the integration time, $T_{\text{sys}}$ denotes the system temperature, $A_{\text{eff}}$ denotes the effective area of each station, $n_p$ denotes the number of polarization, and the $N_{\text{ant}}$ denotes the number of antennas. The sky temperature $T_{\rm sky}$ is the same as the system temperature $T_{\text{sys}}$ of the instrument at this frequency because we assume that our simulation is free from noise. The sky temperature $T_{\text{sky}}= 180\left(\nu/{180\text{ MHz}}\right)^{-2.6}$\,K \cite{Fulanetto2006}. According to the \ref{equ:thermal} equation, the thermal noise is $\sim 4.8$\,$\mu$Jy/beam after $1000$\, hours of integration time with a channel width of $125$\,kHz, an effective area of $962$\,m$^2$ and a system temperature of $333$\,K. This scenario is shown in Figure~\ref{fig:rms_image} by the magenta dashed line for an optimistic scenario. However, instrumental noise was not included in our simulations to isolate the residual ionospheric effects.

\subsection{Tolerances}
\label{imageplaneeffect}
\begin{figure}
    \centering
    \includegraphics[width=\linewidth]{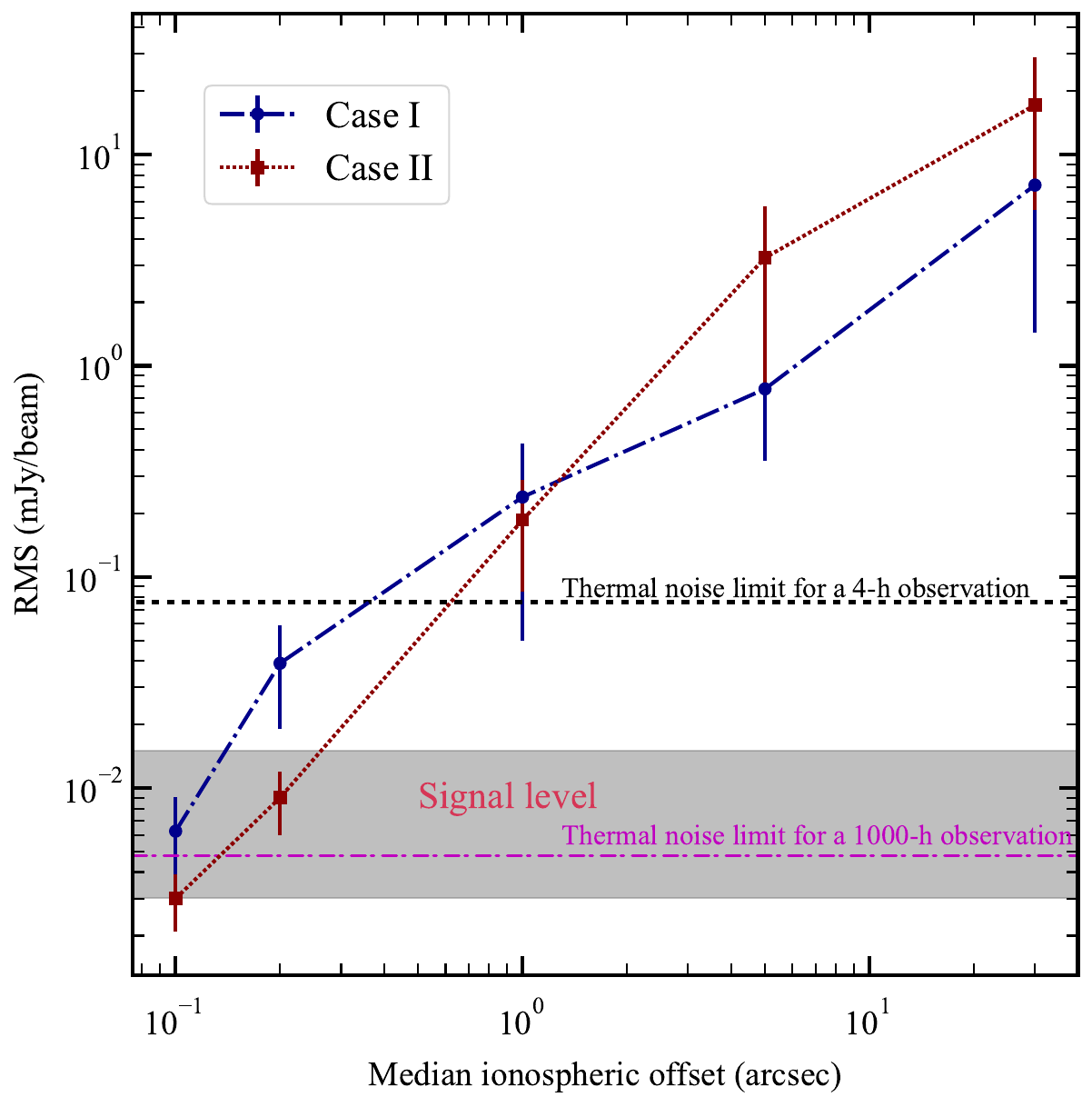}
    \caption{Variation of RMS in the residual image with residual ionospheric errors using T-RECS as foreground model for different ionospheric frameworks considered. The RMS noise levels are plotted against the median ionospheric offset, expressed in arcseconds. The dash-dotted blue line and dotted red line showcase the RMS noise variation in Cases I and II, respectively. The dashed black line and the magenta dashed line represent the theoretical thermal noise limit for a $4$\,h and a $1000$\,h observation time. The grey band represent the \HI\ $21$\,cm signal amplitude observed by the SKA1-Low.} 
    \label{fig:rms_image}
\end{figure}

To quantify the threshold at which residual ionospheric contamination hinders the detection of the \HI\ $21$\,cm signal, we studied the variation in the RMS noise with the MIOs. This variation of RMS noise is observed over a $4$\,hour observation period for different ionospheric frameworks. The comparison of RMS noise variation against MIO of these ionospheric frameworks is shown in Figure~\ref{fig:rms_image} (as discussed in Section \ref{method_imageplane}). The following subsections describe the different cases used in the analysis.

\subsubsection{Case I}
\label{image_plane_ci}
The dash-dotted blue line represents the impact of the residual ionospheric phase errors on the image plane for Case I in Figure~\ref{fig:rms_image}. It is observed that at the MIO value of $0.1''$, the RMS noise falls within the error bars and overlaps with the signal level (represented by the grey band). This represents a clear detection of the target signal in an optimistic scenario, assuming the target signal dominates the observed sky. However, beyond this RMS noise, the residual DDEs suppress the target signal, potentially hindering or obscuring its detection. This highlights that for a sufficient integration time, systematic errors introduced by the {residual DDEs} become more critical than random thermal noise in limiting the observations. In addition, this effect effectively reduces the dynamic range (DR) of the residual image. The observed DR for this scenario is $\sim 10^5$.

\subsubsection{Case II}
\label{image_plane_cii}
The dotted-red line represents the impact of the residual ionospheric phase errors on the image plane for Case II in Figure~\ref{fig:rms_image}. Similarly, at the MIO value of $0.2''$, the RMS noise within the error bar falls in between the target signal level and enables detection of the target signal. The theoretical value of DR is estimated to be $\sim 10^5$, based on Equation \ref{dr_perley}. We achieve a DR of approximately $10^5$ from a dirty image. However, the required dynamic range for the EoR signal detection is approximately $10^8$ \citep{datta_2009}. Therefore, an optimum astrometric accuracy of $0.2''$ is crucial for extracting the \HI\ $21$\,cm signal from the image plane.\\

Figure~\ref{fig:rms_image} also shows the thermal noise limit for $4$\,hours of observation time for SKA1-Low. Datta et al. \citep{Datta_2016} showed that the dynamic ionosphere produces flicker noise (or $1/f$ noise, where $f$ is the dynamical frequency). Hence, ionospheric errors do not integrate down with a longer integration period. There is a need for ionospheric calibration at shorter time intervals matching with ionospheric variability. Here, we assume that the CD/EoR observations will be restricted to $\pm 2$\,hours around the transit time of the target EoR field. Since ionospheric calibration is done over short time scales, we further assume that residual ionospheric errors are not correlated beyond each night/day's observations (i.e. 4\,hours). So, any systematic errors are only restricted within the 4\,hours of observing time. Hence, the RMS noise floor achieved after each epoch (4\,hours) of observations then gets co-added with other epochs, and the RMS noise reduces as $\propto 1/\sqrt{N_{\rm days}}$, where $N_{\rm days}$ is the number of days of observation. Hence, at $1000$\,hours of deep observation period, the desired RMS noise level will be sufficient to extract the $21$\,cm signal in the image domain with higher ionospheric corruptions with $0.2''\lesssim~\theta_{\rm MIO}~\lesssim 1''$. It is important to note that we are not applying any mitigation techniques, as the goal is to estimate the accuracy of any such techniques that will allow us to detect the CD/EoR signal. Therefore, it is essential to develop efficient ionospheric calibration algorithms for the upcoming SKA1-Low observations. A future study will explore the presence and impact of instrumental noise in the extraction of the EoR signal in the image domain in greater detail.

\section{PS estimation}
\label{ps_estimation}
In this section, we outline the method for PS estimation using delay-domain formalization. The impact of residual DDEs from different ionospheric frameworks on the PS estimation is described in the subsequent section.

\subsection{Analysis method}\label{ps_method}
The major aim of a CD/EoR telescope is to detect the PS of a cosmological \HI\ $21$\,cm signal. PS is the estimation of the statistical fluctuation of the \HI\ signal from CD/EoR. The PS can be estimated either directly from the image cube \citep[for more, see][]{Mertens2018MNRAS.478.3640M, Mertens2020} or from the observed visibilities \citep[for more details, see][]{TRott2020, Kern2021, Mazumder2022}; this work employs visibility for its determination.
The inverse Fourier transform of $V(\boldsymbol{u},\nu)$ along the frequency axis gives visibility in the delay domain, which is defined as
\begin{equation}
    V(\boldsymbol{u},\eta) = \int_{\text{sky}} V(\boldsymbol{u},\nu)W(\nu)e^{2\pi i\nu\eta} d\nu
   \label{eq:delayvis}
\end{equation}
where $\eta$ is the time delay and $W(\nu)$ is the window function given by the instrumental FoV and bandwidth. The cylindrical-averaged 2D PS \citep{Morales2004,Thyagarajan2015ApJ...804...14T} is given by
\begin{equation}
    P(k_{\perp},k_{\parallel}) = \left(\frac{\lambda^2}{2k_{B}}\right)^2\left(\frac{A_{\text{eff}}}{\lambda^2 B}\right)\left( \frac{D^2\Delta D }{B}\right)\left|V(\boldsymbol{u},\eta)\right|^2
    \label{2dps}
\end{equation}
In equation \ref{2dps}, $\lambda$ is the wavelength corresponding to the central frequency of the observation, $k_B$ is the Boltzmann constant, $B$ is the bandwidth, $D(z)$ is the transverse co-moving distance and $\Delta D(z)$ is the co-moving depth along the line of sight \citep{Morales2004, Thyagarajan2015ApJ...804...14T} and $A_{\text{eff}}$ is the effective collecting area. 
The wavenumbers $k_{\perp}$  correspond to the sky plane, which is determined by the antenna baseline, and $k_{\parallel}$ represents the line of sight of the sky, which is determined by the spectral resolution. Longer baselines enable the probing of finer angular scales accessible on the sky, corresponding to the largest $k_{\perp}$ modes. The spectral resolution of the interferometer limits the maximum accessible $k_{\parallel}$ modes, while the total bandwidth constrains the minimum accessible $k_{\parallel}$ modes \citep{Liu_2020}. The two-component of the $k$ vector is given below,
\begin{equation}
    k_{\perp} = \frac{2\pi |\boldsymbol{u}|}{D(z)}
    \label{equ:kperp}
\end{equation}
\begin{equation}        
    k_{\parallel} = \frac{2\pi\eta\nu_{\rm rest}H_0E(z)}{c(1+z)^2}
    \label{kpar}
\end{equation}
where $\nu_{\rm rest}$ is the rest-frame frequency of the $21$\,cm spin-ﬂip transition of \HI\ , $z$ is the observed redshift, $H_0$ is the Hubble parameter, and $E(z)\equiv\left[\Omega_{\rm M}(1+z)^3+\Omega_{\Lambda} \right]^{1/2}$. $\Omega_{\rm M}$ and $\Omega_{\Lambda}$ are the matter and dark energy density, respectively, at the present time. The advantage of cylindrical-averaged PS is that it splits the chromatic response of the instrument.
The 1D PS is obtained by the spherical averaging of $P(k_{\perp},k_{\parallel})$ in independent $k$-bins, and the dimensionless 1D PS can be written as
\begin{equation}
    \Delta^2(k)= \frac{k^3}{2\pi^2}P(k)
\end{equation}
where $k=\sqrt{k_{\perp}^2 +k_{\parallel}^2}$. 
To determine the uncertainty in the PS estimation, we added thermal noise to the uncertainty estimation. The noise power spectrum arising from thermal noise contribution in the radio interferometric observation is given by \footnote{This equation is the modified form employed in the literature \cite{Bull2015}}
\begin{equation}
    P_N^{\rm HI} =  \frac{2T^2_{\text{sys}}}{Bt_{\text{obs}}}\frac{D^2(z)\Delta D \Omega_{\text{FoV}}}{n_p}\left(\frac{A_{\text{eff}}A_{\text{core}}}{A^2_{\text{coll}}}\right)
    \label{noise_ps}
\end{equation}
where $A_{\text{core}}$ is the core area of the array, $A_{\text{coll}}$ is the total collection area of the array, $t_{\text{obs}}$ is the total observation hours. We consider the PS uncertainty of the order of $3\sigma$ in a particular $k$-mode. It still needs to be decided with SKA1-Low if foreground removal, avoidance, or the hybrid method is used to estimate the signal PS. The spherically averaged PS was calculated in this work using the general technique of averaging across all $k$-modes. We consider $1000$\,h of observational time to calculate the uncertainty in noise PS. 

\subsection{Systematics effect on PS estimation}
\label{ps_result} 
\subsubsection{Case I}
\label{ps_case1}
\begin{figure*}
    \begin{multicols}{2}
        \includegraphics[width=\linewidth]{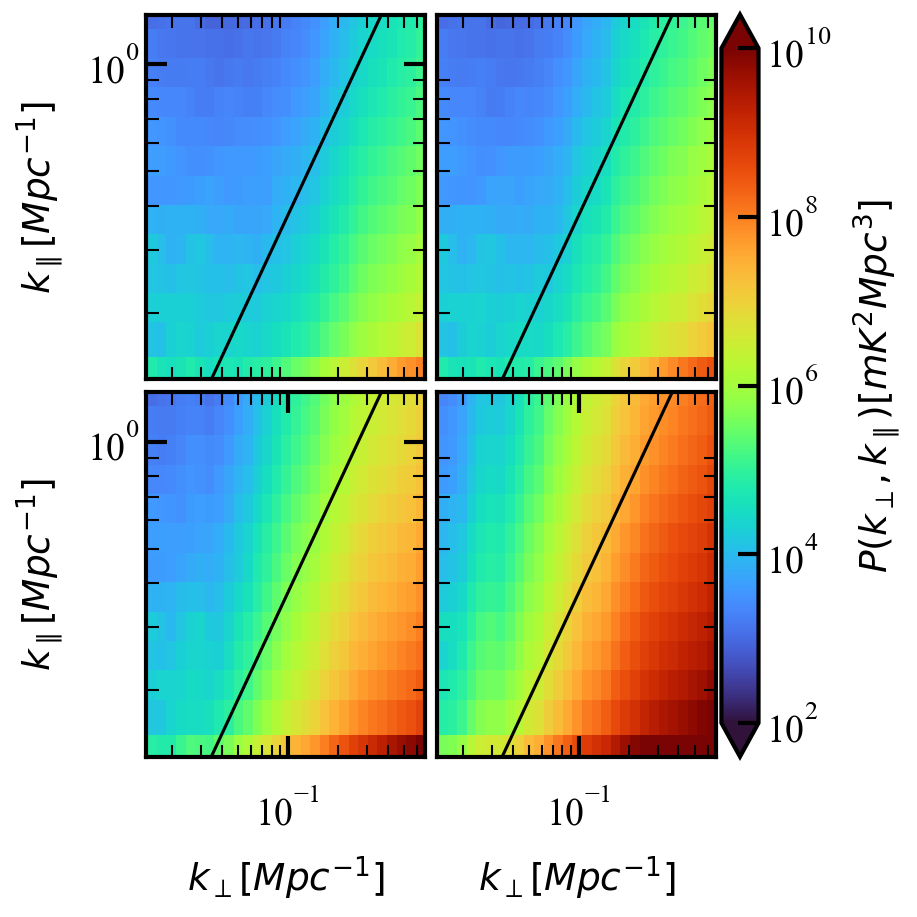}\par
        \includegraphics[width=\linewidth]{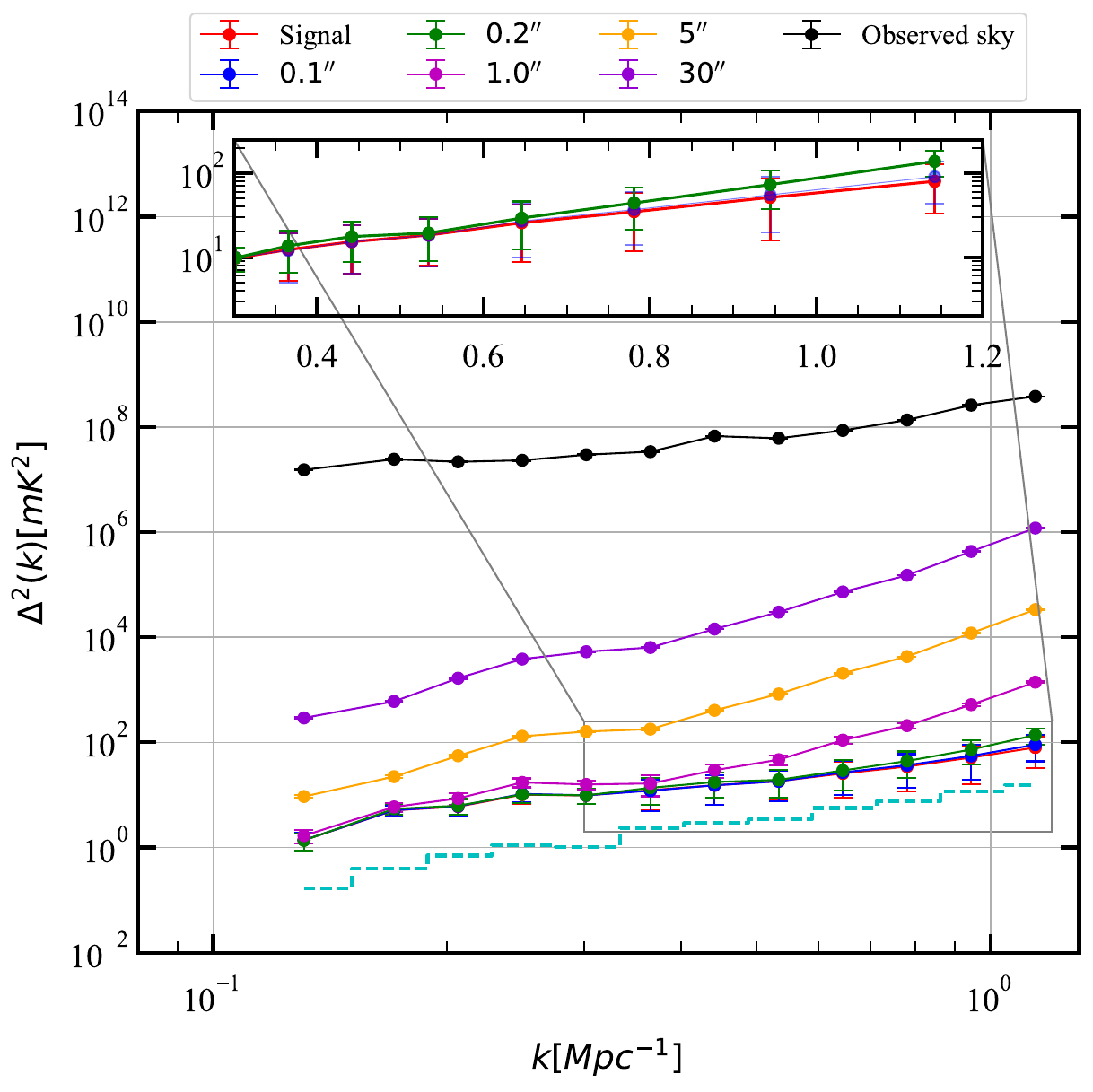}\par
    \end{multicols}
    \caption{Residual PS for direction-dependent calibration errors with SKA1-Low for Case I. \texttt{Left:} Cylindrical averaged power spectra of the residual visibilities for residual ionospheric errors with MIOs of $0.1''$(top-left), $0.2''$(top-right), $1''$(bottom-left), and $5''$(bottom-right). The solid black line represents the horizon line. \texttt{Right:} Comparison of the spherical averaged PS showing residual ionospheric errors with the signal and foreground power. The error bars are the total uncertainties of $3\sigma$  sample variance and thermal noise for all $k$-modes. The cyan line represents the total sensitivity limit of sample variance and $1000$\,hours of observation of instrumental noise.}
    \label{fig:ps_shift}
\end{figure*}

Figure~\ref{fig:ps_shift} presents the PS of residual visibilities for Case I (refer to Section~\ref{ps_method} for details). The left panel shows the cylindrical averaged PS for MIOs of $0.1''$, $0.2''$, $1''$ and $5''$. The solid black line represents the horizon limit. Spectrally smooth foreground interference with ionospheric contamination spills over power beyond the wedge-shaped region in 2D space. Point sources are located at the largest $k_{\perp}$ scales. Imperfect gain calibration leads to residual point sources spilling over power beyond the wedge and into the EoR window in the frequency scales \citep{Bowman_2009}. The foreground contamination becomes more severe when the residual ionospheric offset of sources increases, which signifies active ionospheric conditions. This simulation aims to determine how the accuracy of the ionospheric calibration or the residual ionospheric offsets spill over the residual power out of the wedge and into the EoR window. As expected, the higher ionospheric turbulence level results in greater contamination within the EoR window. The right panel of Figure~\ref{fig:ps_shift} displays the spherical averaged PS of residual visibilities. The solid red and black line represents the \HI\ $21$\,cm signal PS and the observed sky PS (foreground plus \HI\ $21$\,cm signal) as measured by SKA1-Low. The inset highlights the area near the signal power for $k$-modes greater than $0.2$\,Mpc$^{-1}$. The residual PS overlaps with the signal power within the error bars as long as the residual ionospheric shift remains below $\sim 0.2''$. Therefore, the MIO of $0.2''$, for this case, represents the tolerable limit for accurately measuring the signal PS. At the MIO of $1.0''$, the residual amplitude becomes significantly high for all $k$-modes, indicating severe bias in the estimated cosmological PS due to residual foreground contamination.

\subsubsection{Case II}
\begin{figure*}
    \begin{multicols}{2}
    \includegraphics[width=\linewidth]{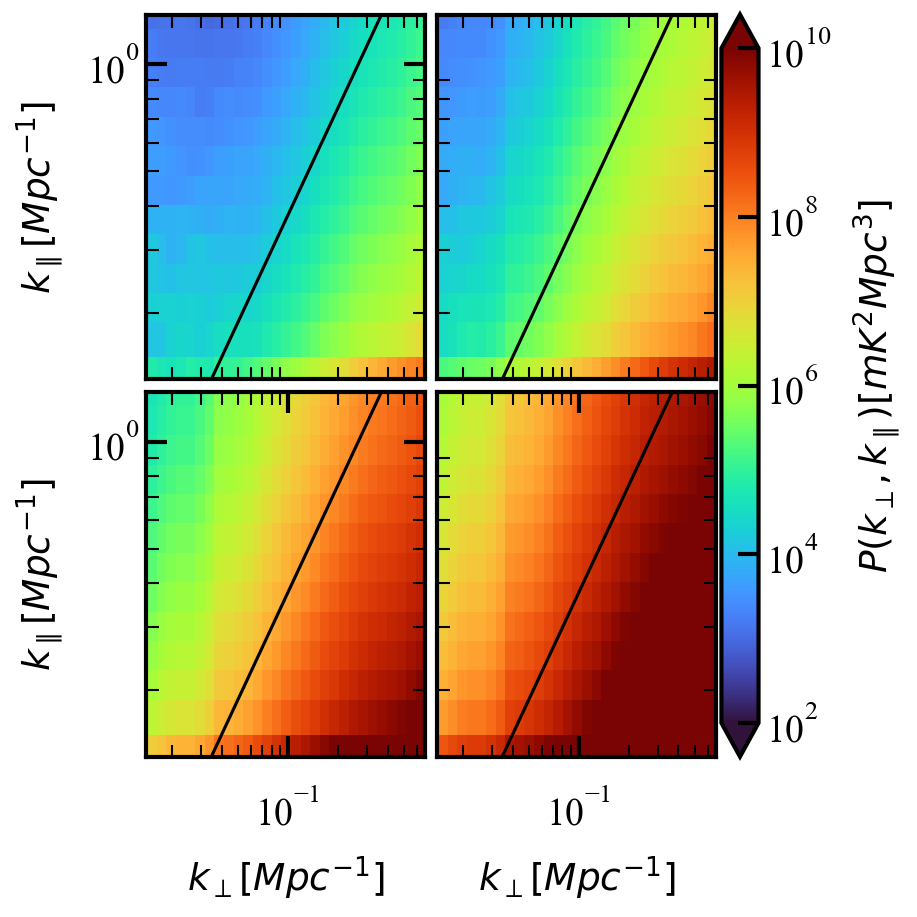}\par
    \includegraphics[width=\linewidth]{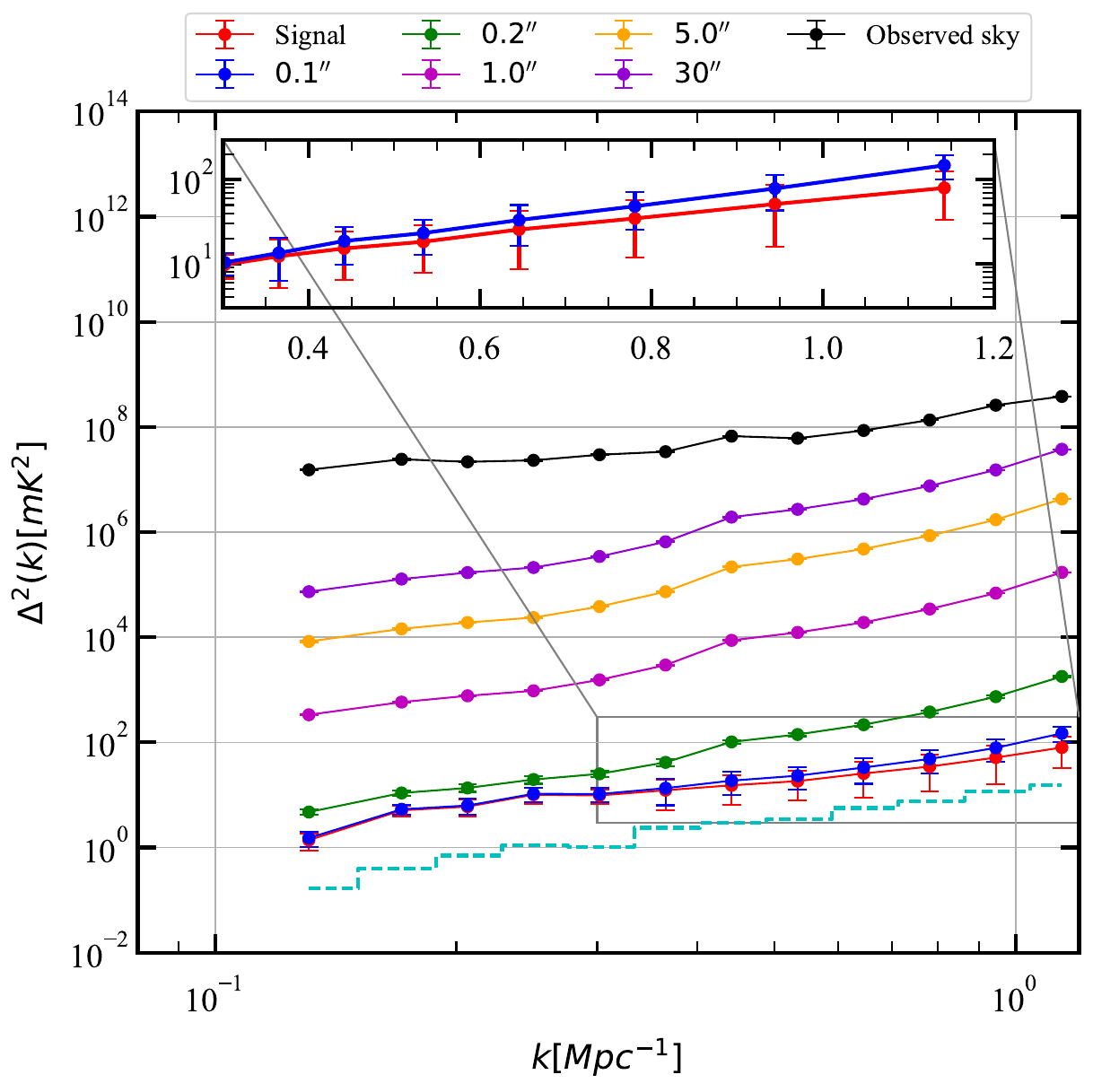}\par
    \end{multicols}
    \caption{Residual PS for direction-dependent calibration errors with SKA1-Low for Case II. \texttt{Left:} Cylindrical averaged power spectra of the residual visibilities for residual ionospheric errors with MIOs of $0.1''$(top-left), $0.2''$(top-right), $1''$(bottom-left), and $5''$(bottom-right).  The solid black line represents the horizon line. \texttt{Right:} Comparison of the spherical averaged PS showing residual ionospheric errors with the signal and foreground power. The error bars are the total uncertainties of $3\sigma$  sample variance and thermal noise for all $k$-modes. The cyan line represents the total sensitivity limit of sample variance and $1000$\,hours of observation of instrumental noise.}
    \label{fig:ps_kolmo}
\end{figure*}

Figure~\ref{fig:ps_kolmo} shows the PS of residual visibilities for Case II under different calibration accuracies with MIOs of $0.1''$, $0.2''$, $1''$ and $5''$. It is observed that the residual DDEs significantly contaminate the EoR as we increase levels of ionospheric calibration inaccuracies,  shown in the left panel of Figure~\ref{fig:ps_kolmo}. This contamination is more pronounced in Case II compared to Case I. As we continue with this figure, the right panel shows the spherical averaged PS using all the accessible $k$-modes. We observed that the residual PS tracks the signal PS for the MIO value of $0.1''$. This indicates that the amount of $0.1''$ calibration accuracy is essential for unbiased estimation of the cosmological PS.

Our finding indicates that to precisely estimate the power spectrum of the $21$\,cm signal originating from the EoR, the calibration accuracy of $\theta_{\text{MIO}} \lesssim 0.1''$ is essential. This initial systematic investigation highlights the importance of implementing ionospheric calibration for the forthcoming SKA1-Low observations. It further emphasizes the necessity of investigating the potential impact of any residual gain errors on the final \HI\ $21$\,cm PS.

\section{Summary and conclusions}
\label{sec:summary}
In this work, we have extended the end-to-end (\textsc{21cmE2E}) pipeline described in \citep{Mazumder2022, Mazumder2023} to investigate the residual DDEs on the recovery of the PS of the \HI\ $21$\,cm signal from the EoR. This target signal has most of its power concentrated at small $k$-scales (or large spatial scales). Thus, for an EoR experiment with an interferometric array, short baselines dictate most of the sensitivity. We study the impact of the residual DDEs using two different frameworks on the shorter baselines in realistic CD/EoR observations using SKA1-Low, focusing on the time-varying nature of the ionosphere. We consider the dynamic ionosphere updates every $2$\,minute cadence with a period of $4$\,hours of observation. However, real-world ionospheric conditions can vary significantly within short timescales, in less than $10$\,seconds, as demonstrated in \citep{Mangla2022, Mangla2023}. The findings of this study can be summarized as follows:
\begin{enumerate} 
    \item Our simulations revealed that different frameworks of ionospheric corruption, including simple phase shifts and turbulence-induced effects, have distinct impacts on the detectability of the target EoR signal. Our analysis suggests that a tolerance level of $0.2''\lesssim ~ \theta_{\rm MIO}~\lesssim 1''$ is necessary to extract the EoR signal in the image domain in the presence of ionospheric corruption. However, this tolerance level should be adapted to account for the variations in RMS noise across different sky fields. When the MIO exceeds this threshold value, residual foreground contamination can significantly hinder the detection of the EoR signal by obscuring it. The required optimum astrometric accuracy ($0.2''\lesssim ~ \theta_{\rm MIO}~\lesssim 1''$) is achievable with upcoming SKA1-Low observations. For example, the LOFAR Two-metre Sky Survey (LoTSS) achieved a position accuracy of $\sim 0.2'' $ between $120$ and $168$\,MHz (\cite{SHIMWELL2019A&A...622A...1S}). The GLEAM survey achieved an accuracy better than $\sim 2''$ between $70$ and $231$\,MHz (\citep{Hurley-Walker2017MNRAS.464.1146H, Hurley-Walker2019PASA...36...47H}). With its improved sensitivity, SKA1-Low is expected to generate sufficiently accurate sky models for making tomographic maps of the EoR.

    \item  At the MRO location, previous studies \citep{Beardsley_2016, jordan2017, Trott2018ApJ, TRott2020, Yoshiura2021, Kariuki2022} have characterized and assessed the impact of ionospheric corruption on the PS estimation. We quantify the level of ionospheric corruption tolerable for precision PS estimation. Our results demonstrate that the recoverable PS of the \HI\ signal dominates any remaining calibration inaccuracies caused by the Earth’s ionosphere depending on the dominant mode of ionospheric corruption. In the case of time-varying phase offsets, with $\theta_{\rm MIO} \sim 0.2''$, the EoR signal remains detectable. However, under the most realistic turbulent conditions modelled by Kolmogorov statistics, residual power becomes excessively high beyond an offset of $\theta_{\rm MIO} \lesssim 0.1''$, thereby hindering \HI\ $21$\,cm signal extraction. The tolerance levels presented here are based on the compact core of SKA1-Low. The impact of ionospheric corruption on longer baselines dictates further investigation and will be a subject of future research. In summary, these results indicate that understanding the ionospheric conditions during observation runs for EoR science is essential because such unaccounted effects, even to the first order, can adversely affect signal extraction. 
\end{enumerate}
This work highlights the critical need for developing efficient ionospheric calibration algorithms for the upcoming SKA1-Low observation. It is to be noted that our current analysis does not account for factors like diffuse foreground emission, thermal noise, and the impact of longer baselines on the observed data. We also assume that other effects like direction-independent gain errors are perfectly accounted for. These omitted complexities further challenge the quantification of ionospheric effects and its overall impact on real data. These considerations are deferred to future studies. We are incorporating further features into the end-to-end pipeline to study more complex systematic effects and test different methods (like Largest Cluster Statistics) to recover the reionization history from observed data in the presence of the real imperfections expected to be present in the SKA1-Low data.

\acknowledgments
SKP acknowledges the financial support by the Department of Science and Technology, Government of India, through the INSPIRE Fellowship [IF200312]. AM thanks the UK Research and Innovation Future Leaders Fellowship for supporting this research [grant MR/V026437/1]. The authors acknowledge the use of facilities procured through the funding via the Department of Science and Technology, Government of India sponsored DST-FIST grant no. SR/FST/PSII/2021/162(C) awarded to the DAASE, IIT Indore. The authors thank the anonymous reviewer and scientific editor for their insightful comments and suggestions, which significantly improved the quality of this manuscript.\\
\textbf{Software}: This work is heavily based on the Python programming language (\url{https://www.python.org/}). The packages used here are \texttt{astropy} (\url{https://www.astropy.org/}, \cite{Astropy2013A&A...558A..33A, Price-whelan2018AJ....156..123A} ), \texttt{numpy} (\url{https://numpy.org/}), , \texttt{h5py} (\url{https://www.h5py.org/}), \texttt{matplotlib} (\url{https://matplotlib.org/}), \texttt{seaborn} (\url{https://seaborn.pydata.org/}), \texttt{scipy} (\url{https://scipy.org/}), \texttt{pyuvdata} (\url{https://github.com/RadioAstronomySoftwareGroup}). \textsc{OSKAR} has been used for simulations (\url{https://github.com/OxfordSKA/OSKAR}), and Common Astronomy Software Applications (\textsc{CASA}) (\url{https://casa.nrao.edu/}) are used for imaging.\\

\appendix
\section{ Impact of constant phase screen}\label{constant_phase_screen}
\begin{figure}
    \centering
    \includegraphics[height= 8cm, width=0.8\linewidth]{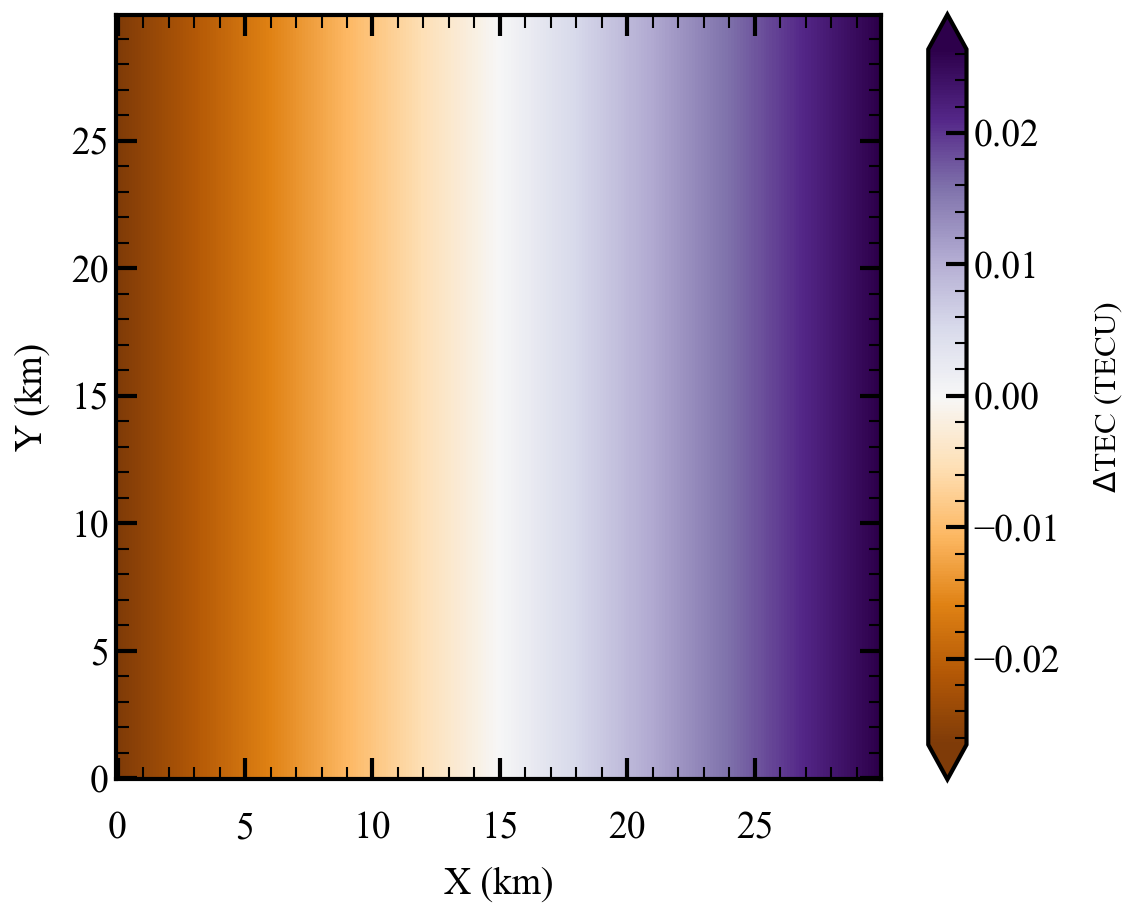}
    \caption{ The two-dimensional phase-screen model has a constant gradient and dimension size of $30$\,km on each side, with $300\times300$ grid size. At the $400$\,km altitude, the ionospheric screen covered the sources within the field of view.}
    \label{fig:tec_grad_phase_screen}
\end{figure}

To test the ionospheric effect on visibility simulations conducted with \textsc{OSKAR}, we inserted a two-dimensional phase screen between the source and the array. This screen represents a simplified ionosphere with a constant phase gradient across the field of view in one direction. In a compact array configuration, all antennas observe a nearly constant gradient of TEC. This constant phase gradient primarily affects the angular position shift of cosmic sources without change distorting the shape of the source. Importantly, Earth aperture synthesis relies on the movement of baseline projections with observation time, effectively filling the u-v plane. Consequently, a static spatial phase gradient transforms into a time-varying gradient in the u-v domain. When we form a synthesized image from a single snapshot observation, this results in source extension and variable amplitude. The phase screen values correspond to a $\Delta$TEC above the station. The OSKAR package \footnote{\url{https://ska-telescope.gitlab.io/sim/oskar/theory/theory.html}} then converts these values into phase shifts and applies them to each source and array element. Figure~\ref{fig:tec_grad_phase_screen} shows the constant gradient ionospheric structure.

\bibliographystyle{JHEP}
\bibliography{myref}




\end{document}